\documentclass[floatfix,showpacs,preprintnumbers,amsmath,amssymb]{revtex4}
\usepackage{graphicx}

\newcommand{\beq}{\begin{equation}}
\newcommand{\eeq}{\end{equation}}
\newcommand{\beqa}{\begin{eqnarray}}
\newcommand{\eeqa}{\end{eqnarray}}
\newcommand{\ba}{\begin{array}}
\newcommand{\ea}{\end{array}}

\newcommand{\etals}{\mbox{\sl et al \/}}

\newcommand{\bra}[1]{\langle #1 |}

\newcommand{\ek}[1]{#1 \rangle}
\begin{document}
\title{R-matrix propagation with adiabatic bases for the
 photoionization spectra of atoms in magnetic fields}

\author{F. Mota-Furtado}
\email{f.motafurtado@rhul.ac.uk}
\author{P. F. O'Mahony}
\email{p.omahony@rhul.ac.uk}

 \affiliation{Department of Mathematics, Royal Holloway, University of London,
Egham, Surrey TW20 0EX, United Kingdom}

\date{\today}

\begin{abstract}

The photoionization spectrum of an atom in a magnetic field is calculated by
combining R-matrix propagation with local adiabatic basis expansions. This approach
considerably increases the speed and the energy range over which calculations can be
performed compared to previous methods, allowing one to obtain accurate partial
and total cross sections over an extended energy range
for an arbitrary magnetic field strength.  In addition, the cross sections for all
atoms of interest can be calculated simultaneously in a single calculation.
Multichannel quantum defect theory allows for
a detailed analysis of the resonance structure in the continuum. Calculated
cross sections for a range of atoms in both laboratory and
astrophysical field strengths are presented.

\end{abstract}

\maketitle

\section{Introduction}

The spectrum of an atom in a magnetic field has played an important
role in the development of quantum theory and in atomic structure \cite{gaz}. For
a long time the effect of the external field on the atom was treated
perturbatively. However with the discovery of large magnetic fields in white
dwarf stars ( $10^2 - 10^5 T$) and neutron stars ( $10^7 - 10^9 T$) in the
seventies attention has focussed on non-perturbative treatments of the field-atom
interaction \cite{rud94}. For laboratory strength magnetic fields ( $ 6 T$)  the potential due to the
applied field only becomes comparable to the intrinsic Coulomb potential
for an electron in a high Rydberg state or continuum state of the atom.
In the eighties and nineties research was focussed
on atoms in laboratory strength fields as these systems
have an inherent non separability arising from the competing
spherical symmetry of the atom and the cylindrical symmetry of the
applied field which leads to the classical system exhibiting chaotic behavior \cite{sch98}.
An atom in a field thus provided an
experimentally realizable quantum system whose corresponding
classical phase space is chaotic, serving as a
prototype for studying classical and quantum chaos.
This lead to fruitful developments in the theory of quantum chaos \cite{gut90}.

For bound states of atoms in a magnetic field
large scale basis set calculations have proved very
successful in finding the energy eigenstates and photoabsorbtion
spectra of atoms in moderately strong fields \cite{rud94}. For super strong fields such as
those found in neutron stars, the field can modify the atomic structure of
the ground state of the atom and different theoretical techniques have to
be used. Currently for these cases the energy eigenstates and photoabsorbtion
spectra are only known for some low levels of a few light atoms \cite{sch04}.

The positive energy or continuum spectra of an atom in a field proved more
challenging particularly in calculating
photoionization cross sections at laboratory strength fields. The three main theoretical methods that
have been successful at calculating cross sections at both laboratory and astrophysical strength
fields are  the complex
coordinate method of Delande \etals \cite{del91}, the R matrix method of
O'Mahony and Mota-Furtado \cite{oma91a,oma91b}, and the diabatic by
sector method of Watanabe and Komine \cite{wat91}.  A detailed comparison between theory and
 experiment has been possible due to the high resolution experiments carried out by
  Iu \etals \cite{iu91} on lithium in a field of about $6T$.
Although all three
approaches have recreated the experimental spectrum of Iu \etals
over a narrow energy region they are not particularly suited to
calculating the photoionization spectrum over a large energy region.
We
present here a major improvement on previous R matrix methods applied to this problem by
using local adiabatic basis states to propagate the R-matrix from
low $r$ to the asymptotic region. This leads to a large saving in both
the CPU time and computer memory required to perform the calculation.
In addition, for a given value of the field strength, the cross sections for all
atoms of interest can be calculated in one step
without the need for any additional propagations.
We demonstrate that by using a combination of R matrix
propagation with local adiabatic basis states and multichannel
quantum defect theory (MQDT) \cite{sea83} one has a method that can be used to calculate
photoionization cross sections of any atom over very large energy regions and
field strengths. MQDT
can also be used to analyze the resonance
structure in detail.
An efficient approach to calculating such cross sections is of importance in many areas where
magnetic fields play a role, for example in  calculating stellar opacities for magnetic white
dwarfs or for evaluating
recombination rates for an atom or ion in a magnetic field at low temperatures (e.g. anti-hydrogen) where one
has to calculate the cross section over very large energy regions for a given field
strength.

In section \ref{sec1} we give the theory used to evaluate the photoionization cross
section of an atom in an external magnetic field by combining  R matrix
propagation with local adiabatic basis states and MQDT.  In section
\ref{sec2} all the required details of the computation are given.  The
results of the calculations are presented in section \ref{sec3} for a variety
of atoms at both laboratory and astrophysical field strengths and a
concluding section is given in section \ref{sec4}. In the Appendix we give details
on how to construct the Hamiltonian matrix for the propagation.

\section{Theory}
\label{sec1}
The Hamiltonian for a hydrogen atom in a magnetic field
(taken to be in the $z$ direction) in the symmetric gauge can be
written using atomic units $ (\hbar = m = e = 1) $ as \cite{gaz},

\beq
\label{eq1.1}
H = -\frac{\nabla^2}{2} - \frac{1}{r} + \beta L_z
+ \frac{1}{2}\beta^2 r^2 \sin^2 \theta
\eeq
where the magnetic field $ B $ is measured in atomic units by $
\beta = B/B_0 $ with $ B_0 = 4.70108 \times 10^5 T $. (We neglect
the spin as it only produces a uniform shift in the energy scale).
The Hamiltonian has two conserved quantities in addition to the
energy, namely the $z$-component of the angular momentum $L_z$ and
$\pi_z$ the $z$-parity. The eigensolutions can thus be studied for
fixed values of $m$ the azimuthal quantum number and for $\pi_z =\pm
1 $.  The linear Zeeman term in eqn. (\ref{eq1.1}) thus only adds a uniform
shift to the total energy.

When a photon excites an electron to the
continuum, for a given magnetic field strength, one can in general
identify three regions of interaction with the continuum electron \cite{oma91b}.
Typically for low $r$ the spherically symmetric Coulomb potential
dominates over the cylindrically symmetric diamagnetic term or
quadratic term in eqn. (\ref{eq1.1}), at intermediate values of $r$ the
Coulomb and magnetic potentials are of comparable strength (the
strong mixing region) and at high values of $r$ or asymptotically in $r$ the
cylindrical symmetry of the diamagnetic potential predominates. For
a general atom (or molecule) one adds a fourth region, the core,
where the excited electron interacts with the multi-electron core
before emerging into the Coulomb region described by the Hamiltonian
in eqn. (\ref{eq1.1}). Exploiting this natural partition in
configuration space forms the basis of the R-matrix approach to
solving atomic and molecular problems where solutions are sought in
each region and then matched together at the boundaries between the
regions to form the solution over all space \cite{bur75}. A novel aspect of the
magnetic field problem is having to deal with the change in symmetry
from spherical to cylindrical which involves introducing
two-dimensional matching procedures \cite{oma91b}. We describe below how the
R-matrix is propagated through the regions described above and how
the two dimensional matching procedure is implemented to give the
reactance matrix and the photoionization cross section.

\subsection{Propagating the R-matrix}

An atom in a magnetic field is assumed to be excited from an initial
state, either a ground or low lying excited state, by a polarized
photon leading to an electron in the continuum with specific
values of $m$ and $\pi_z$. The electron emerges from the first
region, the core region, into the second or Coulomb region with
energy $\epsilon$ where eqn.(\ref{eq1.1}) can be approximated by the
field free Hamiltonian because the diamagnetic terms are negligible
in comparison with those from the Coulomb potential of the atomic
core. (We shall assume here that the field strengths for
non-hydrogenic atoms are not large enough to significantly distort
the core (i.e. $\beta < 1$). In this case a different treatment
would be required for the core although the propagation outside the
core could still be implemented). Therefore at some radius $r =a$ in
the Coulomb region the radial form of the wavefunction, $F_l^{\epsilon} (r)$,
of the
continuum electron can be written in terms of a linear combination
of the energy normalized regular $ s $ and irregular $c$ Coulomb
functions \cite{sea94} in spherical coordinates or a phase shifted Coulomb
function giving the general form of the solution as

\beq
\label{eq1.2}
\Psi_{\epsilon} = \sum_l F_l^{\epsilon} (r) Y_{l
m} (\theta, \phi)= \sum_l A_{l}^{\epsilon} \biggl(
s_{l}^{\epsilon}(r) + c_{l}^{\epsilon}(r) \tan(\pi \mu_l) \biggr) \;
Y_{l m}(\theta,\phi)
\eeq
where the $A_{l}^{\epsilon}$ are constants to be determined. The
quantum defects $ \mu_l$ represent the effects of the non-hydrogenic
field free core \cite{sea83} and can be calculated ab-initio or
taken from experiment. (The more general case of a multi-channel
wavefunction with a reactance matrix instead of quantum defects in
eqn. (\ref{eq1.2}) is straightforward to include). $ Y_{lm}(\theta,\phi) $
are the spherical harmonics. Knowing the phase shifted Coulomb
function and its derivative at $r=a$, the R matrix \cite{bur75} can be
constructed on the outer boundary of this region as

\beq
\label{eq1.3}
R_{ll'}  = \biggl(
s_{l}^{\epsilon}(a) + c_{l}^{\epsilon}(a) \tan(\pi \mu_l) \biggr)\left[ {\left.
{\frac{{d
 }}{{dr}}\biggl(
s_{l}^{\epsilon}(r) + c_{l}^{\epsilon}(r) \tan(\pi \mu_l) \biggr)} \right|_{r = a} }
\right]^{ - 1} \delta _{ll'},
\eeq
since the coefficients $A_{l}^{\epsilon}$ cancel out in the above
expression.

\begin{figure}
\includegraphics[width=0.9\textwidth,height=7cm]{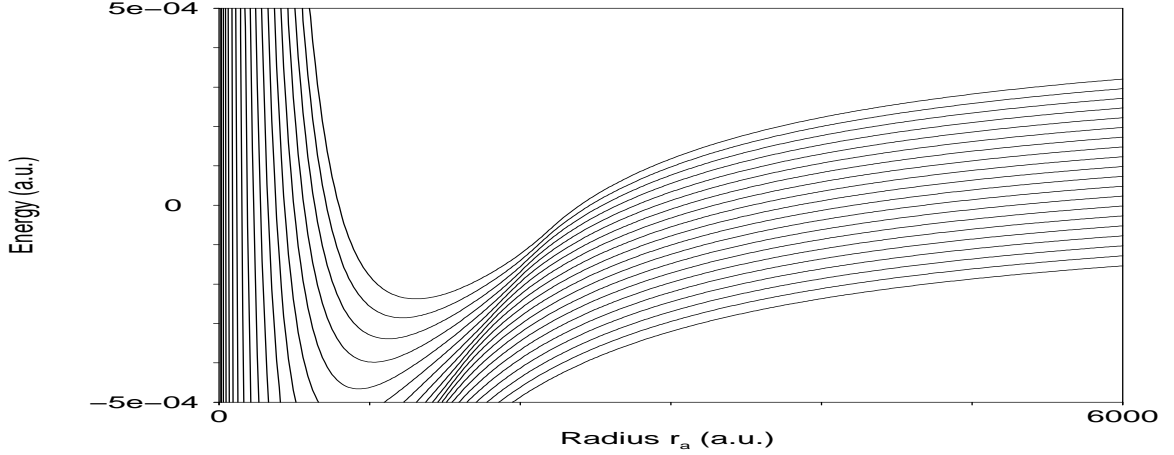}
\caption{\label{fig1} The first 20 adiabatic eigenvalue curves
obtained from diagonalizing the adiabatic Hamiltonian $ H_{a d} $ in
a basis set of spherical harmonics at successive $ r $.  The
magnetic field strength used was $ 6 T $. The centrifugal barrier
can be seem at small $r$ and the Coulomb plus equal Landau spacing
is seen as $ r \rightarrow \infty$.  There is a set of avoided
crossings in between.}
\end{figure}

In the third or strong mixing region the effects due to
the Coulomb potential of the core and those due to the magnetic
field are of a comparable size. This region is defined by $ a < r <
b $ where the radius $b$ is taken to be large enough such that the
Hamiltonian is separable in cylindrical coordinates. The change in
symmetry of the potential, from spherical to cylindrical, is
therefore completely contained within this region. We wish to
propagate the initial R-matrix in eqn. (\ref{eq1.3}) to obtain the R
matrix at the outer boundary of this region $ r=b$ \cite{lig76,bal82}. To do this we
divide the region into $N$ radial sectors with radii $ a \rightarrow
a_1, a_1 \rightarrow a_2, \cdots, a_{N-1} \rightarrow b$. The size
of each sector and the number of sectors $N$ are important
parameters in the calculation and we show how these are optimally
determined later. Within each of the sectors we construct a local
adiabatic basis as follows. For the $n^{th}$ sector we take a radius
$r_a^n$ within the sector $ a_{n-1} < r_a^n < a_n $, which is usually
the mid point of the sector, and we diagonalize the fixed $r$ or
adiabatic Hamiltonian $H_{ad}$
\beq
\label{eq1.4}
 H_{ad} (r_a^n; \theta, \phi) = \frac{{\bf{L^2}}}{2 ({r_a^n})^2}
- \frac{1}{r_a^n}  + \frac12 \beta^2 ({r_a^n})^2 \sin^2 \theta
\eeq
in a basis set of spherical harmonics such that
\beq
\label{eq1.5}
\phi_{\lambda} (r_a^n; \theta, \phi) = \sum_{l}
d_{l\lambda} Y_{l m} (\theta, \phi),
\eeq
where the $d_{l\lambda}$ are constants giving the eigenvalue
equation
\beq
\label{eq1.6}
H_{ad}\phi_{\lambda}=U_{\lambda}(r_a^n)\phi_{\lambda}.
\eeq
The adiabatic states form a locally optimized basis set for each
sector and the adiabatic potential curves can be produced by
plotting the eigenvalues obtained from the diagonalization at
successive $r_a^n$. A typical set of curves is shown in figure
\ref{fig1}. At small $r$ the adiabatic eigenfunctions functions are
spherical harmonics and the potential exhibits the centrifugal barrier.
At large $r$ these curves have the
equal energy spacing of Landau states (combined with the $ -1/r $ fall off)
and the eigenfunctions are localized in the
cylindrical coordinates $\rho$ and $\phi$. The change in symmetry
happens predominantly around a region of avoided crossings that can
clearly be seen in the diagram. The functions $ \phi_{\lambda} $ are
therefore a very good basis
with which to represent the angular part of the wavefunction in the
local region around $ r_a^n $, namely within the sector $n$.

To propagate the R-matrix from sector to sector we use these basis
states within each sector and we diagonalize the full
Hamiltonian (eqn. (\ref{eq1.1})) plus the Bloch operator $L$
 or surface term \cite{blo57},

\beq
\label{eq1.7}
L = \frac12 \biggl( \delta (r-a_n) \frac{d}{d r}
- \delta (r-a_{n-1}) \frac{d}{d r} \biggr) ,
\eeq
in a basis set consisting of a product of orthogonal radial
functions $f_j (r) $ and the adiabatic functions generated for that
sector $\phi_{\lambda} (r_a^n; \theta, \phi)$.
The radial basis functions used are defined in terms of Legendre polynomials
$P_j$ as follows \cite{bal82}

\beq
\label{eq1.8}
f_j (r) = \sqrt{\frac{2 j -1}{a_n-a_{n-1}}} P_{j-1} (u),
\eeq
where $\displaystyle{u = \frac{2}{a_n-a_{n-1}} \left( r - \left(
\frac{a_n+a_{n-1}}{2} \right) \right).}$

The eigenvalue equation is thus
\beq
\label{eq1.9}
(H+L)\Psi^k=E_k \Psi^k .
\eeq
The eigenfunctions obtained from diagonalizing this operator are therefore
\beq
\Psi^k = \sum_{j \lambda} c_{j \lambda}^k \frac{f_j (r)}{r}
\phi_{\lambda} (r_a^n; \theta, \phi).
\label{eq1.10}
\eeq

The total continuum wavefunction at any energy $\epsilon$,
$\Psi_{\epsilon}$,  can be expanded in terms of these R-matrix
eigenstates $\Psi^k$. Since the general solution in the $n^{th}$
sector can also be written as $\Psi_{\epsilon} = \sum_{\lambda}
F_{\lambda} (r) \phi_{\lambda} (r_a^n; \theta, \phi)$ it is
straightforward to show \cite{bal82}, using the operator $H+L$ and eqn. (\ref{eq1.9}),
that the values of the functions and
derivatives on the boundaries of the sector are related by

\beqa
\vec{F}(a_{n-1}) = {\bf r}_2 \vec{F}'(a_n) - {\bf r}_1  \vec{F}' (a_{n-1})  \nonumber \\
\vec{F}(a_n) = {\bf r}_4 \vec{F}'(a_n) - {\bf r}_3 \vec{F}'
(a_{n-1}),
\label{eq1.11}
\eeqa
where the matrix elements of ${\bf r}_1$ to ${\bf r}_4$, called the
sector R-matrices, are given by
\beqa
\label{eq1.12}
( r_{1}^n )_{i j} = \frac12 \sum_{k} \frac{g_{i
k}(a_{n-1}) \; g_{j k} (a_{n-1})} {E_{k}-\epsilon} & & ( r_{2}^n
)_{i j} = \frac12 \sum_{k} \frac{g_{i k}(a_{n-1}) \; g_{j k}(a_n)}
{E_{k}-\epsilon} \nonumber \\
(r_3^n)_{i j} = \frac12 \sum_{k} \frac{g_{i k}(a_n) \; g_{j
k}(a_{n-1})} {E_{k}-\epsilon}& & (r_4^4)_{i j} = \frac12 \sum_{k}
\frac{g_{i k}(a_n) \; g_{j k}(a_n)} {E_{k}-\epsilon},
\eeqa
and
\beq
\label{eq1.13}
g_{\lambda k} (r) = \sum_j c_{j \lambda}^k \frac{f_j (r)}{r} .
\eeq

In summary, knowing the eigenvalues and eigenvectors of eqn.
(\ref{eq1.9}) one can construct the sector R-matrices ${\bf r}_1$ to
${\bf r}_4$  above which, through eqns. (\ref{eq1.11}), relate the
radial solutions and their derivatives on the boundaries of the
sector. The R-matrix relates the function to its derivative, i.e.
$\vec{F}(a_n)={\bf R}(a_n)\vec{F}'(a_n)$, and a simple manipulation
of eqn. (\ref{eq1.11}) yields the relationship between the R-matrix
on the inner and outer boundaries of the sector
\beq
 \label{eq1.14}
 {\bf R}(a_n) = {\bf r}_{4}^n - {\bf r}_{3}^n
\biggl( {\bf r}_{1}^n + \; {\bf R}(a_{n-1}) \biggr)^{-1} {\bf
r}_{2}^n
\eeq
where ${\bf R}(a_n)$ and ${\bf R}(a_{n-1})$ are represented in the
same adiabatic basis set as the sector R matrices.

As the adiabatic basis changes from sector to sector we need finally
to change the basis representation of the R matrix.  The matrix with
elements
\beq
\label{eq1.15}
 ( T^{n-1,n} )_{\lambda \lambda'} = \bra{\phi_{\lambda}
(r_a^{n-1}; \theta, \phi)} \ek{\phi_{\lambda'} (r_a^n ;
\theta,\phi)}
\eeq
is thus constructed.  One uses this transformation to change the basis
representation of R giving
\beq
\label{eq1.16}
{\bf \tilde R} = \overline{{\bf T}}^{n-1,n} {\bf
R} {\bf T}^{n-1,n},
\eeq
where $ \overline{\bf T} $ is the transpose of $ {\bf T}$.

Starting with some initial R matrix it can thus be propagated from sector to sector using
eqns. (\ref{eq1.14}) and (\ref{eq1.16}). Using the
initial input R-matrix given by eqn. (\ref{eq1.3}) the propagation gives
the final R matrix
on the outer boundary in the asymptotic region at $r=b$ with the
final R matrix being represented in the local adiabatic basis of the
last sector.

Although it is possible to propagate the R matrix itself at each of
the sector radii as described above, it is more practical and efficient to derive
global sector R matrices (${\bf R}_1, {\bf R}_2, {\bf R}_3$ and
${\bf R}_4$) relating the first and $n^{th}$ sectors \cite{ste78} which can be
built up sequentially using eqn. (\ref{eq1.11}). One initially generalizes
eqn. (\ref{eq1.11}) to relate the values of the functions and
derivatives on the boundaries of the first and $n^{th}$ sectors
\beqa
\vec{F}(a_{1}) = {\bf R}_2^{n} \vec{F}'(a_n) - {\bf R}_1^{n} \vec{F}' (a_{1})  \nonumber \\
\vec{F}(a_n) = {\bf R}_4^{n} \vec{F}'(a_n) - {\bf R}_3^{n} \vec{F}'
(a_{1}).
\label{eq1.17}
\eeqa
The operator relations for these global sector R matrices, including
the change in basis, have been derived by Stechel \etals \cite{ste78}
and can be obtained by matching the wavefunction and
its derivative on the boundaries between each of the $n$ sectors . They are
\beqa
\label{eq1.18}
 & {\bf R}_1^{n} & = {\bf R}_{1}^{n-1} - {\bf
R}_{2}^{n-1} \overline{{\bf T}}^{n-1,n} {\bf Z}^{n} {\bf T}^{n-1,n}
{\bf R}_{3}^{n-1}
\nonumber \\
& {\bf R}_2^{n} & = {\bf R}_2^{n-1} \overline{{\bf T}}^{n-1,n} {\bf
Z}^{n} {\bf r}_{2}^{n}
\nonumber \\
& {\bf R}_3^{n} & = {\bf r}_3^{n} {\bf Z}^{n} {\bf T}^{n-1,n} {\bf
R}_3^{n-1}
\nonumber \\
& {\bf R}_4^{n} & = {\bf r}_4^{n} - {\bf r}_3^{n} {\bf Z}^{n}
{\bf r}_2^{n} \nonumber \\
{\rm where } \;\; & {\bf Z}^{n} & = \left( {\bf r}_1^{n} + {\bf
T}^{n-1,n} {\bf R}_4^{n} \overline{{\bf T}}^{n-1,n} \right)^{-1}.
\eeqa
In these equations $ {\bf R}_i^n $ are the global sector R matrices
for all sectors up to $n$, ${\bf r}_i^n$ are the sector R matrices
for sector $n$ and ${\bf T}^{n-1,n}$ is the transformation matrix
between the adiabatic basis used in sectors $n-1$ and $n$.
In an analogous way to eqn. (\ref{eq1.14}) the global sector R
matrices at the end of the propagation through all $N$ sectors
are used to relate the R matrix at $r=a$ to the R matrix at
$ r=b $
\beq
\label{eq1.19}
{\bf R}(b) = {\bf R}_4^{N} - {\bf R}_3^{N}
\left( {\bf R}_1^{N} + {\bf R}(a) \right)^{-1} {\bf R}_2^{N}.
\eeq
Note that ${\bf R}_1^{N}$ to ${\bf R}_4^{N}$, determined from
eqn. (\ref{eq1.18}) above, are independent of which
atom one uses so that once they are calculated from the propagation
then ${\bf R}(b)$ for a whole set
of atoms  can be evaluated at once using eqn. (\ref{eq1.19})
by just using the appropriate quantum defects to calculate {\bf R}(a)
in eqn. (\ref{eq1.3}). Hence for given values of $B$, $m$ and $\pi_z$,
the cross sections of all
atoms of interest can be calculated simultaneously without
 the need for any additional propagations.

\subsection{Asymptotic region}
\label{sec2.3}

Having found the R-matrix at $r=b$ by propagation we must match this
to the asymptotic solutions at $r=b$ to find the solution over all
space. For large $ r $, $ r
> b$, the magnetic field dominates. Since the motion in $\rho$ is
bounded $-\frac{1}{r} \rightarrow - \frac{1}{z}$, and the Hamiltonian
in eqn. (\ref{eq1.1}) is separable in cylindrical coordinates
\beq
\label{eq1.20}
H =  -\frac12 \frac{d^2}{d z^2}  - \frac{1}{z} +
H_L + O \left( \frac{1}{z^3} \right)
\eeq
where $H_L$ is the
Hamiltonian for the Landau states
\beq
\label{eq1.21}
H_L  =  - \frac{1}{2}\frac{1}{\rho }\frac{\partial }{{\partial
\rho }}\left( {\rho \frac{\partial }{{\partial \rho }}} \right) +
\frac{{m^2 }}{{2\rho ^2 }} + \beta L_z  + \frac{1}{2}\beta ^2 \rho
^2.
\eeq
$H_L$ has eigenvalues $E_{i}^{L} = (2i+|m|+m+1) \beta, \ \ i=0,1,2
\ldots$ and its eigenfunctions are the Landau states $\Phi_i
(\rho,\phi)$ \cite{gaz}. The asymptotic region $ c \le z \le \infty$, $0 \le
\rho \le \infty $ with $c$ less than the radius $r=b$, is therefore chosen to conform
with the cylindrical symmetry of the problem. In this
region a set of $j$ linearly independent solutions may be written,
as is standard in scattering theory \cite{sea83}, in terms of the solutions of
eqn. (\ref{eq1.20}). These are a product of Landau states $ \Phi_i $
and a linear combination of energy normalized regular and irregular
Coulomb functions in $ z $, $s$ and $c$, evaluated at an energy $
\epsilon_i = \epsilon - E_{i}^{L} $, namely,
\beq
\Psi_{\epsilon j} =
\sum_{i k} \; \Phi_i (\rho,\phi) \; \biggl( s_{i k}^{\epsilon_i} (z)
\delta_{k j} + c_{i k}^{\epsilon_i} (z) K_{k j} \biggr).
\label{eq1.22}
\eeq
The constants $K_{kj}$, the reactance matrix or K matrix,
are to be determined by the matching procedure.

The R matrix ${\bf R}(b) $ having being evaluated using eqn. (\ref{eq1.19}) is
now matched through a two dimensional matching procedure to these
asymptotic solutions on an arc at $ r=b $ \cite{oma91b}. To perform this matching
the integrals
\beq
\label{eq1.23}
\bra{\phi_{\lambda} (b;\theta ,\phi )} \ek{\Psi_{\epsilon j}} =
\int \phi_{\lambda} \Psi_{\epsilon j} \; d \Omega
\eeq
must be evaluated, i.e. the asymptotic solutions must be projected
onto the local adiabatic solutions on the radial arc at $r=b$. This
is done by  evaluating numerically the four matrices ${\bf P}, \; {\bf
Q}, \; {\bf P}', \; {\bf Q}' $ with elements
\beqa
P_{\lambda j} (b)
= \int \; \biggl[ \phi_{\lambda} (b;\theta ,\phi ) \sum_{i} \Phi_{i}
(\rho,\phi) \; s_{i j} (z) \biggr]_{r=b} \;
d \Omega \nonumber \\
Q_{\lambda j} (b) = \int \; \biggl[ \phi_{\lambda} (b;\theta ,\phi )
\sum_{i} \Phi_{i} (\rho,\phi) \; c_{i j} (z) \biggr]_{r=b} \;
d \Omega  \nonumber \\
P_{\lambda j}^{'} (b) = \int \; \biggl[ \phi_{\lambda} (b;\theta
,\phi ) \sum_{i} \Phi_{i} (\rho,\phi) \; s_{i j}^{'} (z)
\biggr]_{r=b} \;
d \Omega \nonumber \\
Q_{\lambda j}^{'} (b) = \int \; \biggl[ \phi_{\lambda} (b;\theta
,\phi ) \sum_{i} \Phi_{i} (\rho,\phi) \; c_{i j}^{'} (z)
\biggr]_{r=b} \; d \Omega
\label{eq1.24}
\eeqa
where $'$ indicates the derivative with respect to $z$. This gives
the regular and irregular components of the solutions at $r=b$ and
allows one to calculate the outer R-matrix from the asymptotic region at
$r=b$ in terms of $K$ the reactance matrix. Equating the inner and
outer R-matrices at $r=b$ gives
\beq
\label{eq1.25}
{\bf R} = \biggl[ {\bf P} + ({\bf Q} \; {\bf K}) \biggr]
\biggl[ {\bf P}' + ({\bf Q}' \;{\bf K}) \biggr]^{-1} .
\eeq
Re-arranging this expression gives us the equation for the K matrix
\beq
\label{eq1.26}
{\bf K} = \biggl[ ({\bf R} \; {\bf Q}') - {\bf Q} \biggr]^{-1}
\biggl[ ({\bf R} \; {\bf P}' ) - {\bf P} \biggr].
\eeq
Knowing ${\bf K}$ we have the energy normalized solution over all
space and we can calculate both partial and total photoionization
cross sections.

\subsubsection{Multichannel quantum defect theory}
\label{sec1.6}

For a given total energy $\epsilon$ there are two possibilities for
the behaviour of the Coulomb solutions in eqn. (\ref{eq1.22}). If $
\epsilon_i = \epsilon- E_{i}^{L} > 0 $ the channel is open and the Coulomb functions
oscillate at $r=b$ and all the way to infinity. If $ \epsilon_i <
0 $ then the channel is closed and the solutions must decay as $z
\to \infty$.  The physical ${\bf K}$ matrix therefore is a square
matrix with the dimension of the number of open channels. However
with multichannel quantum defect theory one exploits the known
analytic properties of the Coulomb functions to enforce the boundary
conditions for the closed channels.  For the closed channels there
are two possible scenarios at $r=b$. Either the channel is strongly
closed and is already exponentially small at $r=b$ or it is weakly
closed, i.e. the Coulomb functions are still oscillating at $r=b$
before decaying at infinity. In MQDT, for a weakly closed channel, one
instead uses Coulomb functions in eqn. (\ref{eq1.22}) which don't
decay at infinity and one treats the channel as if it is open only
enforcing the boundary condition at infinity analytically in a final
step \cite{sea83}. Therefore the resonance structure due to these weakly closed
channels can be calculated analytically.  Hence the K matrix
calculated by doing the matching using MQDT, denoted by ${\bf \cal K}$,
has dimension given by the number of open plus weakly closed
channels.
In general $ \cal{K}$ has a smooth dependence on energy because the
Rydberg series of resonances converging on the Landau thresholds
corresponding to the weakly closed channels have not as yet been included.

The open part of the actual physical reactance matrix $K$ can be recovered
from the matrix $ \cal{K}$ by the
formula \cite{sea83}
\beq
{\bf K}_{o o} = {\bf \cal K}_{o o} - {\bf \cal K}_{o c}
\biggl( \tan ( \pi \nu ) + {\bf \cal K}_{c c} \biggr)^{-1} {\bf \cal
K}_{c o}.
\label{eq1.27}
\eeq
The $o$ and $ c $ subscripts refer to
the open and closed channels of ${\cal K}$ and the $ \tan \pi \nu 's
$ form a diagonal matrix where the $\nu$'s are related to the
energies $ \epsilon_i$ by
\beq
\label{eq1.28}
\nu_i = \frac{1}{\sqrt{2 \mid \!
\epsilon_i \! \mid}}.
\eeq

This way of obtaining the $K$ matrix  has two major advantages.
Firstly it is much quicker computationally since $ {\cal K}$ represents
a reactance matrix with a lot of the resonance structure removed it
varies much more slowly with energy than the full reactance matrix $ K $.  This
allows one to calculate $ { \cal K} $ on a coarse energy mesh with
fairly large energy spacings.  These $ { \cal K} $'s can then be used to
calculate $K$ over an arbitrarily fine energy mesh using the analytic
formula in eqn. (\ref{eq1.27}). This allows the propagation stage to be
performed at fewer energy points ultimately speeding up the
calculation enormously.  The second major advantage of this approach is that
the resonance structure converging to a particular Landau threshold can be
identified by removing it from $K$. This is done by keeping open the
relevant Landau channel in the evaluation of equation (\ref{eq1.27}).
By comparing a spectrum with all resonances converging on a
particular Landau threshold removed with that of the full spectrum
it is possible to determine which resonances converge to which
thresholds \cite{oma91b,wan91}.  This technique is demonstrated in section \ref{sec3}.

\subsection{Photoionization cross section}
\label{sec2.5}

The photoionization cross section is given by
\beq
\label{eq1.29}
\sigma = 4 {\pi}^2 \alpha \omega |\bra{\Psi_{\epsilon}^{-}
}{ \Vec{\epsilon} . \Vec{r}}| \ek{\Psi_{o}}|^2
\eeq
where $\alpha$ is the fine structure constant, $\omega$ the photon
energy, $\Vec{\epsilon}$ the polarization direction, $\Psi_{o}$ the
initial bound state and $\Psi_{\epsilon}^{-}$ the energy normalized
'incoming' wavefunction \cite{sob}. There is a standard transformation to go
from the K-matrix form given in (\ref{eq1.22}) to the S-matrix or
incoming form $\Psi_{\epsilon}^{-}$ \cite{sea83}. Once the $K$ matrix and hence the
asymptotic form is
known from the matching, the photoionization cross section is
evaluated by calculating the amplitude of the wavefunction near the
origin, and hence the dipole integrals, as follows.

The 'incoming' wavefunction on the inner boundary of the
strong mixing region ($r=a$) can be written as
\beq
\label{eq1.30}
\Psi_{\epsilon}^{-}
= \sum_l F_l^{-} (a) Y_{l m} (\theta, \phi).
\eeq
The wavefunction
on the outer boundary of the strong mixing region can be written as
\beq
\label{eq1.31}
\Psi_{\epsilon}^{-} = \sum_{\lambda} G_{\lambda}^{-} (b)
\phi_{\lambda} (b; \theta, \phi),
\eeq
where $G_{\lambda}^{-}$ is
the energy normalized 'incoming' asymptotic solution calculated from the
matching procedure in section \ref{sec2.3}.   The radial solution $ \vec{F}^{-} $ and $
\vec{G}^{-} $ are linked by the global sector R matrices as given in eqn. (\ref{eq1.17})
so that
\beq
\label{eq1.32}
\vec{F}^{-}(a) = {\bf R}_2^{N} \vec{G}'^{-}(b) - {\bf R}_1^{N}
\vec{F}'^{-} (a).
\eeq
The radial solutions on $r=a$
can be written as $ \vec{F}^{-} = { \bf S} \vec{A}^{-} $ from eqn.
(\ref{eq1.2}) where {\bf S} is a diagonal matrix with elements
\beq
\label{eq1.33}
s_l^{\epsilon} (a) + c_l^{\epsilon} (a) \tan \pi \mu_l
\eeq
and
$\vec{A}^{-}$ are the field and energy dependent amplitudes.
Substitution into equation eqn. (\ref{eq1.32}) yields the equation
\beq
\label{eq1.34}
({\bf S} + {\bf R}_1^{N} {\bf S}' ) \vec{A}^{-} = {\bf R}_2^{N}
\vec{G}'^{-}(b).
\eeq
The coefficients $\vec{A}^{-}$ can therefore
be evaluated by solving the set of linear equations because $\vec{G}
'^{-} (b) $ is known once the $K$ matrix is known. Using the
coefficients $\vec{A}^{-}$ one can then simply express the cross
section in terms of the field free photoionization cross section as
the coefficients $\vec{A}^{-}$ give the difference in amplitudes
between the field dependent and field free amplitudes . For example
for excitation from the $1 s $ state of hydrogen using linear
polarized light ($\Delta m = 0 $) only the $l=1$ component of the
final state will be accessed.  The photoionization cross section in
this case is therefore given by
\beq
\label{eq1.35}
\sigma (\epsilon) = \mid \!
A_1^{-}(\epsilon) \! \mid^2 \sigma_{B=0}.
\eeq
where $ \sigma_{B=0}
$ is the field free photoionization cross section for hydrogen.

\section{Computational details}
\label{sec2}

\begin{figure}
\includegraphics[width=0.9\textwidth,height=8cm]{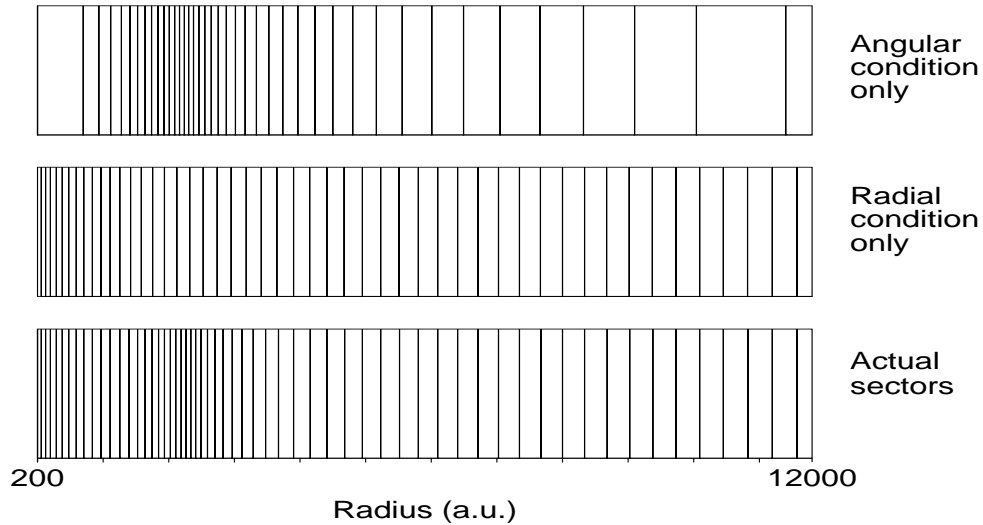}
\caption{\label{fig2} The sector radii obtained using the angular
condition only, the radial condition only and both conditions as
described in the text.  The first and last sector radii are 200 atomic
units and 12000 atomic units respectively.  The magnetic field
strength is $ 6 T $.}
\end{figure}

The first point to address is the choice of the radii $a$ and $b$.
The inner radius must be taken larger than the core ($ \sim 1$ $a.u.$)
for non-hydrogenic atoms yet small enough that the diamagnetic term
is still negligible compared to $ -1/r$. For example we took $a=200$
atomic units for laboratory strength fields. For the asymptotic
radius $b$, extensive calculations show that it is surprisingly
large indicating that the long range coupling due to the terms of
$O(1/z^3)$ falls off very slowly. Based on our experience we used the
following empirical formula for an arbitrary field
\beq
\label{eq1.36}
b = 700\left( {\frac{{10^{ - 3} }}{\beta }}
\right)^{2/3} .
\eeq
The number of sectors between $r=a$ and $r=b$ and their sizes are
chosen in the following way. First one chooses the maximum energy to
be used in the calculation, $\epsilon_{max}$.
 The transformation matrix $( T^{n-1,n} )_{\lambda \lambda'}$
  between adiabatic functions evaluated at two
radii $r_a^{n-1} $ and $ r_a^n $ for consecutive sectors can be
calculated using eqn. (\ref{eq1.15}).  If $r_a^{n-1} $ was equal to
$ r_a^n $, $( T^{n-1,n} )_{\lambda \lambda'}$ would be the identity
matrix. As the distance $ \mid \! r_a^{n-1} - r_a^n \! \mid $ increases the
corresponding adiabatic functions become more different.  The effect
this has on $ T $ is that the diagonal elements get smaller and the
off diagonal elements get bigger.  The maximum sector size for the
angular adiabatic basis is found by choosing a limit on how small any
of the diagonal elements of $ T $ can become. Choosing a value of
$0.5$ for the smallest diagonal element restricts the size of each
sector for a laboratory strength field to those shown in figure
\ref{fig2}. One can see that for small and large $r$ the adiabatic
functions don't vary much with $r$ being close to spherical
harmonics and Landau states respectively leading to large sector
sizes. The intermediate range of $r$, where the avoided crossings in
the potential curves shown in figure \ref{fig1} are present, is
where the angular functions are changing rapidly and one requires
small sectors.

The radial basis sector widths are chosen by comparison with
the local wavelength of the Coulomb functions, namely
\beq
\label{eq1.37}
{\rm
Radial \; sector \; size} = \frac{{\rm constant}}{\left[ 2\left(
\epsilon_{max} + \frac{1}{r}\right) \right]^{\frac12}}.
\eeq
If a radial basis of ten Legendre polynomials is taken, it is found
that a $ {\rm constant} = 6.0 $ is sufficient to produce an
accurate description of the wavefunction \cite{bal82}. For the maximum energy
used, the sectors resulting from this criterion are also shown in
figure \ref{fig2}.
There are therefore two criterion for choosing the
sector sizes; one from the radial basis and one from the angular
basis. The smallest value of these two determines the actual sector
sizes and these are also shown in figure \ref{fig2}.

The adiabatic functions $\phi_{\lambda} ( r_a^n ; \theta,\phi)$
are obtained by diagonalizing the adiabatic Hamiltonian $
H_{a d} $ in a basis of spherical harmonics.  The number of
spherical harmonics used must ensure the functions $ \phi_{\lambda}
( r_a^n ; \theta,\phi) $ are properly converged.  At the matching
radius $ r=b $, the ratio between the diamagnetic term and the
potential term in $ H_{a d} $  will be at its greatest. It is
therefore at this radius that the largest number of spherical
harmonics will be required because it is here that the adiabatic
functions will have their greatest degree of cylindrical symmetry.
We choose the number of $ l$'s so that the
eigenvalue corresponding to the second closed channel at $r=b$ for an energy $
\epsilon_{max} $   is accurate to $ 0.5 \% $.  It has been verified
that this is sufficient to give convergence in the final cross
section.

When the R matrix is propagated from $ r \! = \! a \rightarrow r \!
= \! b $, it is necessary to retain, within any one sector, all of
the locally open channels plus a few of the locally closed ones. The
number of closed channels required is directly related to the
threshold value on the adiabatic angular functions used to determine
the sector sizes. The number of closed channels needed, however, is
constant for all of the sectors.   For a given energy $\epsilon$,
because the number of locally open channels changes with radius (see
figure \ref{fig1}), the total number of channels retained in any one
sector varies. Because it is very difficult to include a channel
half way through the propagation, the number of channels retained in
a sector is found by the following method.  Firstly the maximum
number of open channels retained  for a given $\epsilon_{max}$ in
any one sector is found and this sector is labeled $ k $.  In every
sector up to and including $ k $ the number of channels retained is
this maximum number of open channels plus the number of extra closed
channels. After sector $ k $ the number of channels retained is the
number of locally open channels plus the number of closed channels.
The redundant channels in the global sector R matrices are removed
for any given energy $\epsilon$ by a simple truncation as it is
required during the propagation stage. Essentially once the saddle
point of the adiabatic curves (see figure \ref{fig1}) is passed the
number of channels retained in the R matrix propagation can be
reduced. An example of the number of channels retained in a program
run for a magnetic field strength $ 470 T $ is shown in table
\ref{tab1} showing that from about $190$ a.u. on the number of
channels propagated decreases with $r$.

\begin{table}[htbp]
\begin{center}
\begin{tabular}{|c|c|c|}
\hline
Outer sector radius & no. of local open channels & no. of channels used \\
\hline \hline
23.41 & 3 & 8 \\
43.94 & 3 & 8 \\
72.05 & 4 & 8 \\
108.04& 5 & 8 \\
141.02& 6 & 8 \\
189.01& 4 & 6 \\
247.23& 3 & 5 \\
313.78& 2 & 4 \\
388.78& 1 & 3 \\
472.01& 1 & 3 \\
563.75& 1 & 3 \\
663.93& 1 & 3 \\
700.00& 1 & 3 \\
\hline
\end{tabular}
\caption{Table of the number of open channels and the total number of
channels used in each sector of the propagation for a magnetic field
strength of $ 470 T $.  Two extra closed channels are retained in
each sector.} \label{tab1}
\end{center}
\end{table}

The number of channels needed to match the R matrix to asymptotic
solutions in order to obtain the $ {\cal K}$  matrix was checked for each
spectrum to ensure convergence.
Once the $ {\cal K}$  matrix was obtained, however, the number of
channels needed for the calculation of the cross section was reduced
to the number of open channels plus the number of weakly closed
channels.  Weakly closed channels were taken to be those that were
within an energy (in $a.u.$) of $ \frac{1}{b} $ of their
corresponding Landau energy.

The cross section is obtained over a coarse mesh of energies
initially and then MQDT is used to calculate the cross section over
an arbitrarily small energy range.
The convergence of the cross sections was tested by varying the
inner and outer radii , $a$ and $b$, and the other parameters in the
calculations.

\section{Photoionization cross sections for laboratory strength fields}
\label{sec3}

\begin{figure}
\centerline{\includegraphics[width=0.5\textwidth,height=16cm,angle=-90]{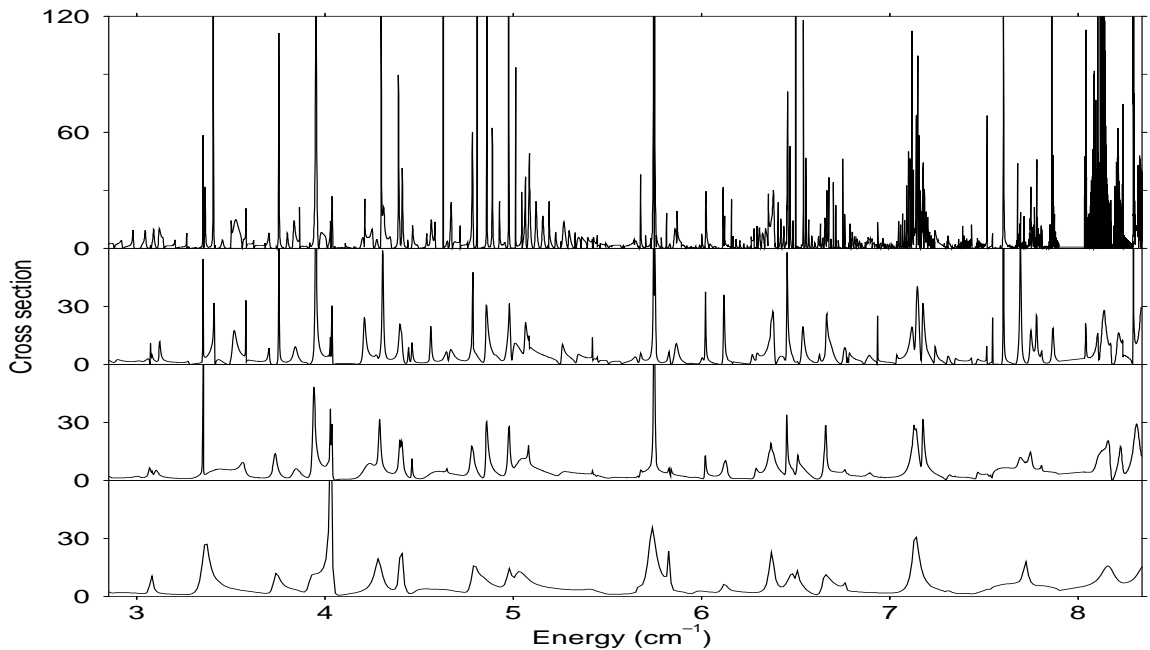}}
\caption{\label{fig3} Photoionization cross section of lithium in
arbitrary units versus the energy in wavenumbers measured relative to the field
free ionization threshold. Excitation is from the $3s$ state in a
field strength of $6.1143 T$ using linearly polarized light. The
final state has $m=0$ and $\pi_z = -1$. The top panel shows the full
spectrum and the second, third and fourth panels show the spectrum
with all the resonances converging on the nearest, second nearest
and third nearest threshold removed using multichannel quantum
defect theory.}
\end{figure}

Calculations at laboratory strength fields are the most demanding
numerically as the radial distances over which one needs to propagate the
R-matrix are large. The number of channels needed in the calculation can
also become quite large. We focus on the photionization spectrum of
lithium in a magnetic field of $ 6.1143 T $ ($ \beta = 1.3 \times
10^{-5} $) where experimental data exist (Iu et al \cite{iu91}). Lithium is
excited from the $3s$ state using linearly polarized light giving a
final state with $m=0$ and $\pi_z = -1$ or odd z-parity.  We
 show in figure \ref{fig3}
 the calculated spectrum between the photoionization threshold (i.e.
$i=0$ or first Landau level) and the second Landau level or
threshold $i=1$.

The only significant quantum defect is for $l=1$ , $\mu_p =0.053$.
(The $l=0$ continuum state doesn't play any role for excitation from
$s$ states.) The photoionization cross section is given in arbitrary
units but it can be put on an absolute scale if the field free cross
section is known. The radius $a$ is $200$ and $b=12600$. The
adiabatic matrix threshold was taken to be 0.1. Together with an
$\epsilon_{max}$ of $ 3.9 \times 10^{-5} $ a.u. this gives the number of
sectors to be 64. The maximum number of locally open adiabatic
channels is 27 and taking 13 extra closed channels the maximum
number of channels overall is 40, hence the maximum size matrix to
be diagonalized is 400 since there are 10 radial basis functions per
sector. The cross section is calculated over a course mesh at $400$
energy points between the thresholds. The full spectrum is then
calculated semi-analytically using MQDT and eqn. (\ref{eq1.27})
with over 10,000 energy points.
The spectrum obtained from this calculation is displayed in the top
panel of figure \ref{fig3} showing the complete resonance structure
between the thresholds. MQDT can also be used to examine the
resonance structure by keeping individual weakly closed channels
open, i.e. by not enforcing the closed channel conditions
asymptotically. For a single threshold this is equivalent to a
Gailitis average \cite{gal63,sea83} over the resonances converging to that threshold.
For several thresholds this is a type of generalized Gailitis
average over the resonance structure and would be similar to
convoluting the actual spectrum with a Gaussian of a certain width \cite{wan91}.  The
differences between the first and second panel allows one to
identify the perturbed resonances which converge to the second
Landau level or threshold $i=1$. The third panel in figure \ref{fig3} has
resonances converging on the next two thresholds removed ($i=1$ and
$i=2$) and the final panel has resonances converging to the $i=3$ threshold
also removed. Some of the remaining modulations may be resonances converging on
higher Landau levels but cannot be removed because the
wavefunctions of these resonances are completely contained
within the radius $ b =12600 $. A lower matching radius would be
required to further reduce the spectrum. Note that the final panel
contains only about 20 resonances compared to the very large number
in the first panel indicating that much of the complex resonance
structure is due to several Rydberg series interacting with a finite
number of short range perturbers.

\begin{figure}
\centerline{\includegraphics[height=9cm]{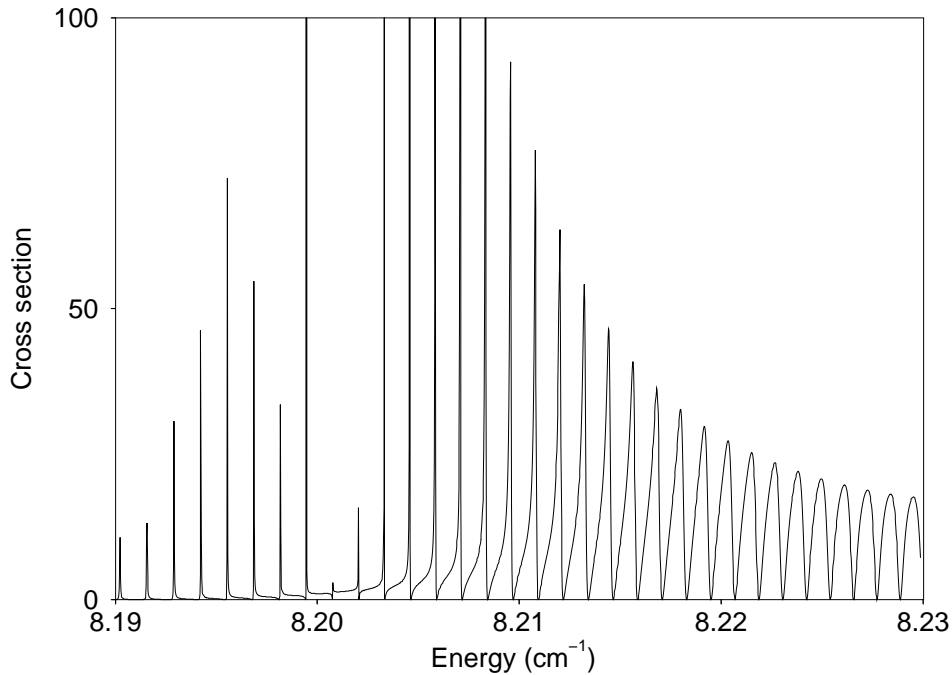}}
\caption{\label{fig4} Photoionization cross section of lithium, in
arbitrary units, over a very small energy range just below the first
excited Landau threshold. This cross section demonstrates the
arbitrary resolution of the theoretical technique.}
\end{figure}

In addition to helping in the analysis of the spectrum,  MQDT  allows
one to calculate the resonance structure to an arbitrarily small
resolution. This is demonstrated in figure \ref{fig4} where an
enlargement of the spectrum in a very small energy region just below
the first excited Landau level is shown.

\begin{figure}
\centerline{\includegraphics[width=0.5\textwidth,height=16cm,angle=-90]{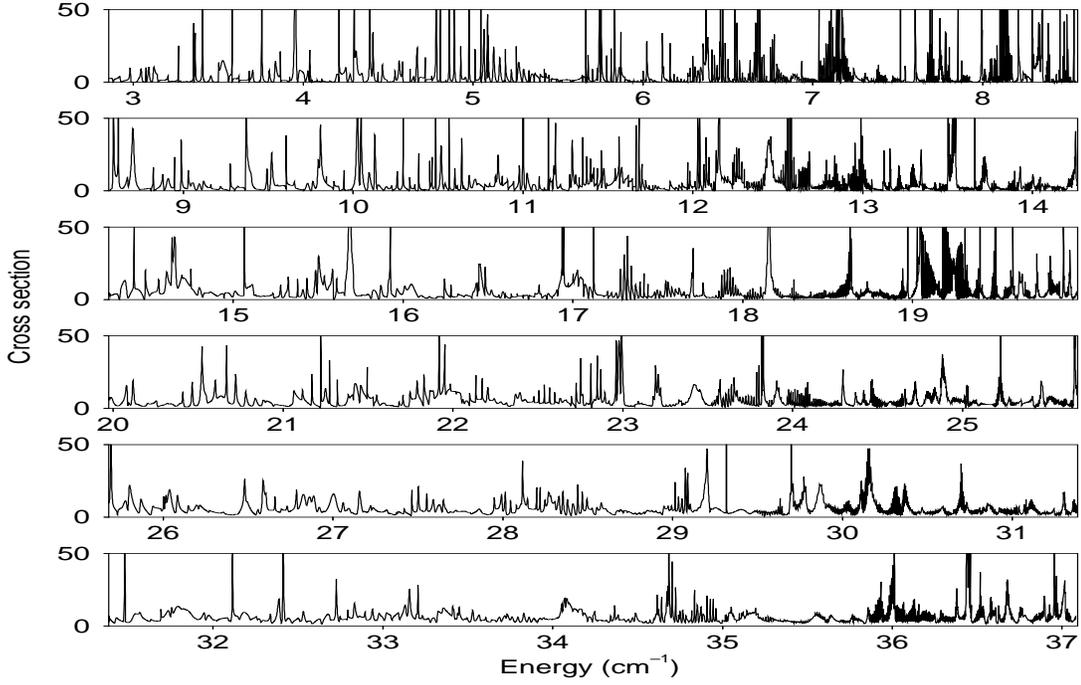}}
\caption{\label{fig5} The extension of the photoionization spectrum
of lithium in a magnetic field of $ 6.1143 T $ shown in figure \ref{fig3}
to an energy range covering over 6 Landau thresholds.  Each panel
shows the spectrum over one Landau threshold.}
\end{figure}
However the real power of the combination of R matrix propagation
with adiabatic bases can be seen in figure \ref{fig5}, where we calculate
the photoionization cross section over an  extended energy region
covering over  6 Landau thresholds from the ionization threshold.
Because the number of open channels increases with energy the size
of the matrices to be diagonalized are slightly larger. The total
number of sectors increases also but even for the highest energy,
the number of channels didn't exceed $50$ giving the largest
matrices that need to be diagonalized to be of the order of 500. The
cross section was calculated on a coarse energy grid of 500 points
over each of the thresholds before applying MQDT. The method scales
in a reasonable way thus allowing one to calculate the
photoionization cross section of an atom over a large energy range
above the ionization threshold.

\section{Photoionization cross sections for astrophysical strength fields}
\label{sec4}

\begin{figure}
\centerline{\includegraphics[height=8cm]{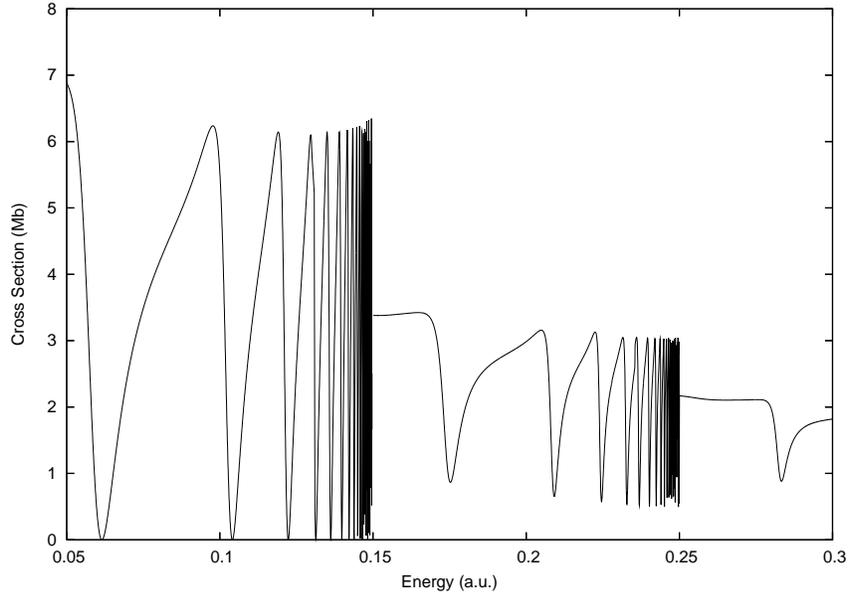}}
\caption{\label{fig6} Photoionization cross section of hydrogen in
megabarns versus the energy in a.u. measured relative to the field
free ionization threshold. Excitation is from the $1s$ state in a
field strength of $23,500T$ using linearly polarized light. The
final state is $m=0$ and $\pi_z = -1$. The energy range covers the
first couple of Landau thresholds. }
\end{figure}

Photoexcitation and photoionization cross sections of light elements
such as hydrogen and helium are important in understanding the
properties of white dwarf and neutron stars \cite{rud94}. The method detailed in
section \ref{sec1} can
equally well be applied to atoms in astrophysical strength magnetic
fields. In fact the computation is much easier in this case compared
to laboratory strength fields as the radius $b$ is much smaller and
the number of channels and sectors is smaller too. We give just two
examples to illustrate the point. We first consider the
photoionization spectrum of hydrogen from the ground state using
linearly polarized light in a field of 23,500T ($\beta = 0.05$ a.u.)
(See figure \ref{fig6}.) The radius $a$ was $1$ a.u. and $b=50$, the number of
sectors used was $20$ and up to ten channels were used in the propagation. This
spectrum has also been calculated using the complex co-ordinate
method \cite{del91,zha06}. Excellent agreement is found between the two methods away
from the energy region near to the ionization thresholds. As the complex co-ordinate method uses a
finite basis it cannot represent all of the Rydberg structure just below the ionization thresholds.

\begin{figure}
\centerline{\includegraphics[height=8cm]{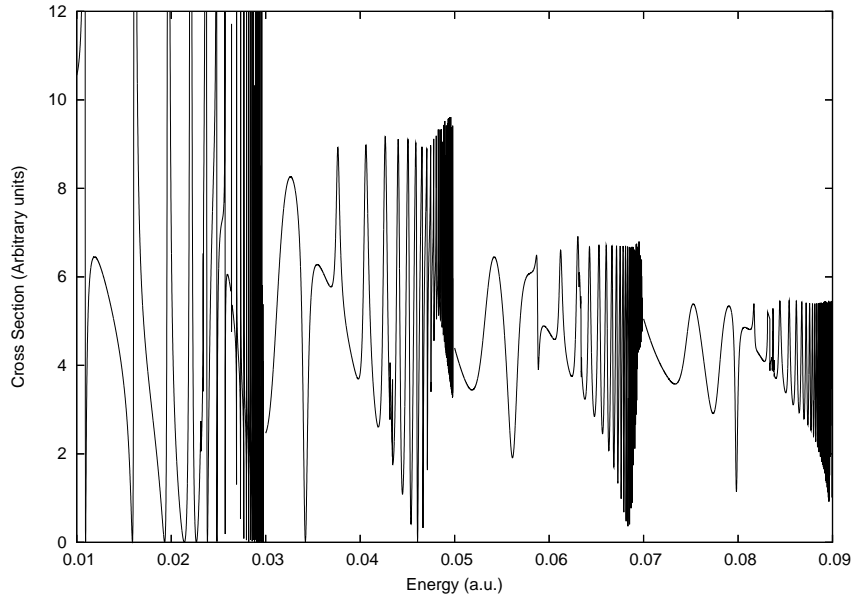}}
\caption{\label{fig7} Photoionization cross section of helium in
arbitrary units versus the energy measured relative to the field
free ionization threshold. Excitation is from the ground state in a
field strength of $4700T$ using linearly polarized light.}
\end{figure}

The second example is the photoionization spectrum of helium from
the ground state in a field of 4700T ($\beta = 0.0001$ a.u.).  The
only significant quantum defect is for $l=1$, $\mu_p =-0.012$. (The
$l=0$ continuum state will only play a role for excitation from $1s2p$
state.) The radius $a$ was $5$ a.u. and $b=150$, the number of
sectors used was $20$ and a maximum of
twelve channels were used in the propagation. Resonances
converging to the first 4 excited Landau thresholds are shown in
figure \ref{fig7}. The resonance structure is that of strongly perturbed
Rydberg resonances converging to the individual thresholds.

The spectra in both cases, when calculated over an extended energy range, repeat the
patterns shown in figures \ref{fig6} and \ref{fig7}.

\section{Conclusions}
\label{sec5}

We have presented a detailed description of a method to evaluate the photoionization
cross section of an atom in an external magnetic field. By combining R-matrix propagation
with local adiabatic bases we have shown that is possible to calculate
the cross section over an extended energy range for a range of field strengths.
We have calculated cross sections for a range of atoms in both laboratory and
astrophysical field strengths to
illustrate the generality of the method.
In addition, for given values of $B$, $m$ and $\pi_z$, the spectra of all
atoms of interest can be calculated without the need for any additional propagations.
By using MQDT one is able to calculate quickly all of the resonances in the
spectrum and to analyze some of their main characteristics.
Partial cross sections to individual Landau levels are evaluated when
calculating the total cross section enabling one to calculate their distributions.
The method can be used to calculate the large amounts of data needed for such
problems as stellar opacities or for calculating re-combination rates at low temperatures for an
atom in a magnetic field.

\section{Appendix}
In constructing the matrix representation of the Hamiltonian plus Bloch operator
in each sector one has to construct matrices for the operators
\beq
\label{eq1.38}
-\frac12 \frac{d^2}{d r^2} +
\frac{l(l+1)}{2 r^2} - \frac{1}{r}
\eeq
and $r^2$ with the radial basis set of shifted Legendre polynomials.

For an arbitrary sector with inner radius $r=a$ and outer radius $r=b$ the
coordinate $ r $ can be rescaled to the coordinate $ u $ such that
\beq
\label{eq1.39}
u = \frac{2}{b-a} \left( r - \left( \frac{b+a}{2} \right)
\right).
\eeq
The limits of the integrals in $ r $ over the limited
range $ a \rightarrow b $ therefore become $ -1 \rightarrow 1 $ in
the coordinate $u$.  The orthogonal radial functions are thus
\beq
\label{eq1.40}
f_n (r) = \sqrt{\frac{2 n -1}{b-a}} P_{n-1} (u).
\eeq
where $P_n$ is the Legendre polynomial of order $n$.
The kinetic term can be evaluated by integration by parts and using the relation (Copson
\cite{cop} page 282)
\beqa
\label{eq1.41}
\int_{-1}^{+1} \frac{ d P_{n}(u)}{d u}
\frac{ d P_{m} (u) }{d u} d u & = &
n(n+1) \; \; \; {\rm if} \; \; m-n \; \; {\rm even} \nonumber \\
&=& 0 \hspace{1.6cm} \; {\rm if} \; \; m-n \; \; {\rm odd}
\eeqa
for
$ m \ge n $,
hence
\beqa
\label{eq1.42}
\frac12 \int_a^b \frac{d f_n}{d r} \frac{d
f_m}{d r} d r
& = & \frac12 \frac{\sqrt{(2 n-1 )( 2 m-1)}}{(b-a)^2} n ( n-1) \; \;
\;
{\rm if } \; \; m-n \; \; {\rm even} \nonumber \\
& = & 0 \hspace{5.4cm} \; {\rm if} \; \; m-n \; \; {\rm odd}
\eeqa
where $ m \ge n $.

The potential term $I_{nm}$ is given by
\beq
\label{eq1.43}
I_{nm} = \int_a^b f_n (r)
\left( - \frac{1}{r} \right) f_m (r) d r = \frac{\sqrt{(2 n-1 )( 2
m-1)}}{b-a} (-1) \int_{-1}^{+1} \frac{P_{n-1} (u) P_{m-1} (u)} {u +
\frac{b+a}{b-a} } d u.
\eeq
Using the relation
\beq
\label{eq1.44}
\int_{-1}^{+1}
\frac{P_n (x) P_m (x)}{z-x} d x = 2 P_n (z) Q_m (z)
\eeq
where $ m \ge n $ and $ Q_m (z) $
is a Legendre function of
the second kind \cite{cop} the potential term becomes
\beq
\label{eq1.45}
I_{nm} =
\frac{\sqrt{(2 n-1 )( 2 m-1)}}{b-a} 2 P_{n-1} \left( - \frac{
(b+a)}{b-a} \right) Q_{m-1} \left( - \frac{(b+a)}{b-a} \right).
\eeq
Since
\beqa
\label{eq1.46}
P_n ( -z) = (-1)^n P_n (z) \nonumber \\
Q_n (-z) = (-1) ^{n+1} Q_n (z)
\eeqa
one gets finally
\beq
\label{eq1.47}
I_{nm} = \frac{\sqrt{(2 n-1 )( 2 m-1)}}{b-a}
2 (-1)^{n+m+1} P_{n-1} \left( \frac{ b+a}{b-a} \right) Q_{m-1}
\left( \frac{b+a}{b-a} \right).
\eeq
The centrifugal
term requires the integrals $J_{nm}$
\beq
\label{eq1.48}
J_{nm} = \int_a^b f_n (r) \frac{1}
{r^2} f_m (r) d r = \frac{ \sqrt{(2 n -1 ) ( 2 m - 1)}}{ (b-a)^2}
\int_{-1}^{+1} \frac{P_{n-1} (u) P_{m-1} (u)}{\left( u + \frac{b+a}{b-a}\right)^2} d
u .
\eeq
The integral in this equation can be evaluated by
differentiating eqn. (\ref{eq1.44}) to obtain
\beqa
\label{eq1.49}
\int_{-1}^{+1}
\frac{P_n (x) P_m (x)}{(z-x)^2}
& = & \frac{-2}{z^2 - 1} \biggl( (n+1) P_{n+1} (z) Q_m (z) + (m+1)
P_n (z)
Q_{m+1} (z) + \nonumber \\
& & (n+m+2) z P_n ( z) Q_m (z) \biggr)
\eeqa
where the following
relations have been used
\beqa
\label{eq1.50}
(z^2 - 1) \frac{d P_n (z)}{d z}& =
&(n+1) P_{n+1} (z) - (n+1) z P_n (z)
\nonumber \\
(z^2 - 1) \frac{d Q_m (z)}{d z}& = &(m+1) Q_{m+1} (z) - (m+1) z Q_m
(z).
\eeqa
Substituting the appropriate value for $z$ and using the relations
in eqn. (\ref{eq1.46}) the final
result is
\beqa
\label{eq1.51}
J_{nm} = \frac{ \sqrt{(2 n -1 ) ( 2 m -
1)}}{ (b-a)^2} \frac{(-2)}{c^2 - 1} (-1)^{n+m} \biggl( n P_n (c)
Q_{m-1} (c) + \nonumber \\ m P_{n-1} (c) Q_m (c) - c (n+m) P_{n-1}
(c) Q_{m-1}(c) \biggr),
\eeqa
where
\beq
\label{eq1.52}
c = \frac{b+a}{b-a}.
\eeq

To evaluate the integral involving $ r^2 $, the recurrence
relation
\beq
\label{eq1.53}
(2 n+1) u P_n (u) = (n+1) P_{n+1} (u) + n P_{n-1} (u)
\eeq
is used.  This yields
\beqa
\label{eq1.54}
\int_a^b f_n (r)
r^2 f_m (r) d r = \frac{ (b-a)^2}{4} \biggl( \frac{ n
(n+1)}{(2 n + 1) \sqrt{2 n-1} \sqrt{2 n - 3}}
\delta_{m n+2}  \hspace{1.5cm} \nonumber \\
+\frac{2 c n}{\sqrt{2 n -1} \sqrt{2 n +1}} \delta_{m n+1} + \left(
c^2 + \frac{n^2}{(2 n-1) (2 n + 1)} + \frac{(n-1)^2}{(2 n-1) (2
n-3)}
\right) \delta_{n m} \nonumber \\
+ \frac{2 c (n-1)}{\sqrt{2 n-1} \sqrt{2 n -3} } \delta_{m n-1} +
\frac{(n-1) (n-2)}{(2 n-3) \sqrt{2 n-1} \sqrt{2 n - 5} } \delta_{m
n-2} \biggr). \hspace{2.7cm}
\eeqa

To evaluate equations eqn. (\ref{eq1.47}) and eqn. (\ref{eq1.51}) the functions $
P_n(c) $ and $ Q_n(c) $ must be calculated. To calculate the Legendre
polynomials $ P_n (c) $ the standard recurrence
relation
\beq
\label{eq1.55}
P_{n+1} (c) = \frac{(2 n + 1)c}{n+1} P_n (c) -
\frac{n}{n+1} P_{n-1} (c)
\eeq
is used where the
first two Legendre polynomials are given by $ P_0 (c) = 1 $ and $
P_1 (c) = c $.
The method required to calculate $ Q_n(c) $ depends on the value $ c $.
For $ c < 1 $ the same
recurrence relation as eqn. (\ref{eq1.55}) can be used
 with $ Q_0 (c) = \frac12 \ln \left( \frac{c+1}{c-1}
\right) $ and $ Q_1 (c) = \frac{c}{2} \ln \left( \frac{c+1}{c-1}
\right) - 1 $.  For the case $ c = \frac{b+a}{b-a} > 1 $ a
different method is used.  For $ c
> 1 $ the recurrence relation in  eqn. (\ref{eq1.55}) should only be
used for decreasing values of $n$.  The first two values can be
evaluated using the expression
\beq
\label{eq1.56}
Q_n(c) = \frac12 P_n (c) \ln
\left( \frac{1+c}{1-c} \right) - W_{n-1} (c),
\eeq
where
\beqa
\label{eq1.57}
W_{n-1} (c) & = & \frac{2 n -1}{1 (n)} P_{n-1} (c) + \frac{ 2 n -
5}{3 (n-1)}
P_{n-3} (c) + \frac{2 n - 9}{5 (n-2)} P_{n-5} (c) + \cdots \nonumber \\
& = & \sum_{m=1}^n \frac{1}{m} P_{m-1} (x) P_{n-m} (x).
\eeqa
The first values could also be calculated
using hyper-geometric functions via
\beqa Q_n(c) =
\label{eq1.58}
\frac{\pi^\frac12}{2^{n+1}}
\frac{\Gamma(n+1)}{\Gamma(n+\frac{3}{2})} \frac{1}{c^{n+1}} F \left(
1 + \frac{n}{2} , \frac12 + \frac{n}{2} ; n + \frac{3}{2} ;
\frac{1}{c^2} \right)
\eeqa
for $ \mid \! c \! \mid > 1 $, however
in the calculations described in this paper the expression in
eqn. (\ref{eq1.56}) was used.  The Legendre functions of the second kind
can therefore be evaluated using the recurrence relation
\beq
\label{eq1.59}
Q_{n}
(c) = \frac{n+2}{n+1} Q_{n+2}(c) - \frac{2 n + 3}{n+1} c Q_{n+1}
(c).
\eeq

\end{document}